\documentclass[fleqn,usenatbib]{mnras}

\usepackage{newtxtext,newtxmath}

\usepackage[T1]{fontenc}

\DeclareRobustCommand{\VAN}[3]{#2}
\let\VANthebibliography\thebibliography
\def\thebibliography{\DeclareRobustCommand{\VAN}[3]{##3}\VANthebibliography}

\usepackage{graphicx}	
\usepackage{amsmath}	

\newcommand{\kms}{\,km\,s$^{-1}$}   
\newcommand{\Msun}{M$_{\sun}$}     
\newcommand{\Lsun}{L$_{\sun}$}     
\newcommand{\Msuny}{$M_{\sun}$\,yr$^{-1}$}      
\newcommand{\Hal}{H$\,{\small \alpha}\,$}     
\newcommand{\DNa}{Na\,{\sc i}  D \,}     
\newcommand{\Na}{Na\,{\sc i}\,}     
\newcommand{\Teff}{$T_{\textrm{eff}}$~}     
\newcommand{\vsini}{$v \sin i$}     

\title[Modulated accretion in RY Tau]{Modulated accretion in T Tauri star RY Tau  --  stable MHD propeller or a planet at 0.2 AU?}

\author[P. P. Petrov et al.]{
P. P. Petrov,$^{1}$\thanks{E-mail: petrov@craocrimea.ru}
M. M. Romanova,$^{2}$
K. N. Grankin,$^{1}$
S. A. Artemenko,$^{1}$
E. V. Babina,$^{1}$
\newauthor {S. Yu. Gorda$^{3}$}
\\
$^{1}$Crimean Astrophysical Observatory, p/o Nauchny,  298409, Republic of Crimea\\
$^{2}$Cornell University, Ithaca, NY 14853, USA\\
$^{3}$Ural Federal University, 51, Lenin av., Ekaterinburg, Russia, 620000
}

\date{Accepted 2021 March 26. Received 2021 March 24; in original form 2021 January 08}

\pubyear{2021}

\begin{document}
\label{firstpage}
\pagerange{\pageref{firstpage}--\pageref{lastpage}}
\maketitle

\begin{abstract}
Planets are thought to form at the early stage of stellar evolution when the mass accretion is still ongoing. RY Tau is a T Tauri type star at the age of a few Myr, with accretion disc seen at high inclination, so that line of sight crosses both the wind and the accretion gas flows. In a long series of spectroscopic monitoring of the star in 2013-2020, we detected variations in \Hal and \DNa absorptions at radial velocities of infall (accretion) and outflow (wind) with a period of  about 22 days. The absorptions in the infalling and the outflowing gas streams vary in anti-phase: an increase of infall is accompanied by a decrease of outflow, and vice versa. These `flip-flop' oscillations retain phase over several years of observations.
We suggest that this may result from the MHD processes at the disk-magnetosphere boundary in the propeller mode.
Another possibility is that a massive planet modulates some processes in the disc and provides the observed effects. 
The period, if Keplerian, corresponds to a distance of 0.2 AU, which is close to the dust sublimation radius in this star.  The presence of the putative planet may be confirmed by radial velocity measurements: expected amplitude is $\geq 90$ \,m\,s$^{-1}$ if a planet mass is $\geq 2 $M$_J$.
\end{abstract}

\begin{keywords}
Stars: variables: T Tauri, Herbig Ae/Be -- Stars: winds, outflows -- Line: profiles -- Stars: individuals: RY Tau
\end{keywords}



\section{Introduction}

Accretion plays a major role in the observed properties of Classical T Tauri stars (CTTS) -- low mass (< 2-3 \Msun) stars at the early stage of their evolution \citep[see reviews by][]{ Bouvier2007, Hartmann2016}. CTTS host accretion discs, where planetary systems formation is going on. In recent years, the ALMA interferometer provides new evidence of the planetary formation processes around CTTS \citep*[e.g.][]{Dong2018}.

The observed activity of CTTS, including their rich emission line spectrum and irregular light variability, are due to magnetospheric accretion, winds outflow, and the dusty circumstellar environment. Accretion is a driving force of most of these processes. Both accretion and wind flows are nonstationary, which makes the investigation challenging. 
Magnetospheric accretion works in different astrophysical objects: neutron stars, white dwarfs in cataclysmic variables, and  young stars. 
Accretion in young stars is relatively slow, as compared to compact objects. Therefore, a CTTS is a convenient laboratory to study the accretion in detail. In parallel to observations, numerical simulations of the accretion and winds in a variety of astrophysical objects have made significant progress \citep[see review by][]{Romanova2015}.

From spectroscopic observations, the direction and velocity of the gas flows around CTTS can be traced from emission and absorption line profiles broadened by the Doppler effect. 
In the visual spectrum, \Hal line profile is the main indicator of wind, while the red-shifted absorptions formed in the accretion funnels are most visible in  \DNa and the O\,{\sc i} 7773 triplet lines. A reach sample of the inversed P Cyg profiles of metal lines was observed in CTTS S CrA SE \citep{Petrov2014}.
A combination of spectroscopy and photometry provides information about the flux variability in emission lines.

In the Crimean Astrophysical Observatory (CrAO) we carry out a long monitoring program of two CTTS: RY Tau and SU Aur. The aim is to follow the dynamics of accretion and wind flows on time-scales from days to years. A major part of the program was published in \citet{Petrov2019},  hereafter referred to as Paper I.
Now we present new results, including the seasons of 2018-2019 and 2019-2020, with emphasis on the discovery of modulated accretion and wind flows around RY Tau.

RY Tau is one of the brightest and most studied CTTS. With stellar mass M = 2.08 \Msun, luminosity L = 13.5 \Lsun and \Teff = 5945 K
(Paper I) 
the star drops in between CTTS and more massive Herbig Ae-Be stars (HAEBE). The mass accretion rate, estimated from UV luminosity,  is $6.4 - 9.1$ in units of $10^{-8}$ \Msuny \citep{Calvet2004}. From ALMA survey of protoplanetary discs around young stars \citep{Long2019}, 
the accretion disc of RY Tau is seen at high inclination,  $i = 65\pm 0.1^\circ $, 
so that line of sight to the star intersects the wind flows, the dusty atmosphere of the inner disc, and the accretion funnels within the magnetosphere. Irregular variability of brightness is mostly due to the circumstellar extinction \citep*{Babina2016}. Evidence of stellar occultation by the disc atmosphere close to the sublimation rim at 0.2 AU from the star was received from K-band interferometry \citep{Davies2020}.

No rotational modulation of brightness was detected with a certain confidence in extended photometric series of RY Tau, although some quasi-periods were reported \citep[e.g.][]{ Herbst1987, Zajtseva2010}. From the stellar parameters and the observed \vsini $= 52 \pm 2$\kms \citep{Bouvier1990, Petrov1999}, and assuming the inclination $i = 65^\circ$ (the same as in accretion disc), the period of the axial rotation should be about 3 days, i.e. RY Tau belongs to fast rotators. RY Tau has an extended jet with a few knots of young dynamical ages \citep{Onge2008}. The jet is wiggled, which was interpreted as due to the presence of an unseen planetary or sub-stellar companion to the star \citep{Garufi2019}.

Interferometric image of the protoplanetary disc around RY Tau at millimeter wavelengths did not reveal planets more massive than 5 M$_J$ at distances between 10 and 60 AU \citep*{Isella2010}.

This paper is organized as follows. First, we address briefly the instruments and observation sites and outline the amount of data collected. Then we analyse the variability of \Hal and \DNa lines in terms of accretion and wind flows, with emphasis on the `flip-flop' effect. Finally, we discuss the results in terms of the propeller regime with the possible presence of a planet.

\section{Observations}

Our program of spectral and photometric monitoring of RY Tau has been underway since 2013. Description of the telescopes and instruments at different observatories, used in this monitoring, is given in Paper I.
The major part of the spectroscopic data was obtained at CrAO with an echelle spectrograph at 2.6 meters Shajn reflector. The registered spectral regions included \Hal and \DNa lines, with resolution $\lambda/\Delta\lambda = 27000$.

\begin{figure}  
	\includegraphics[width=\columnwidth]{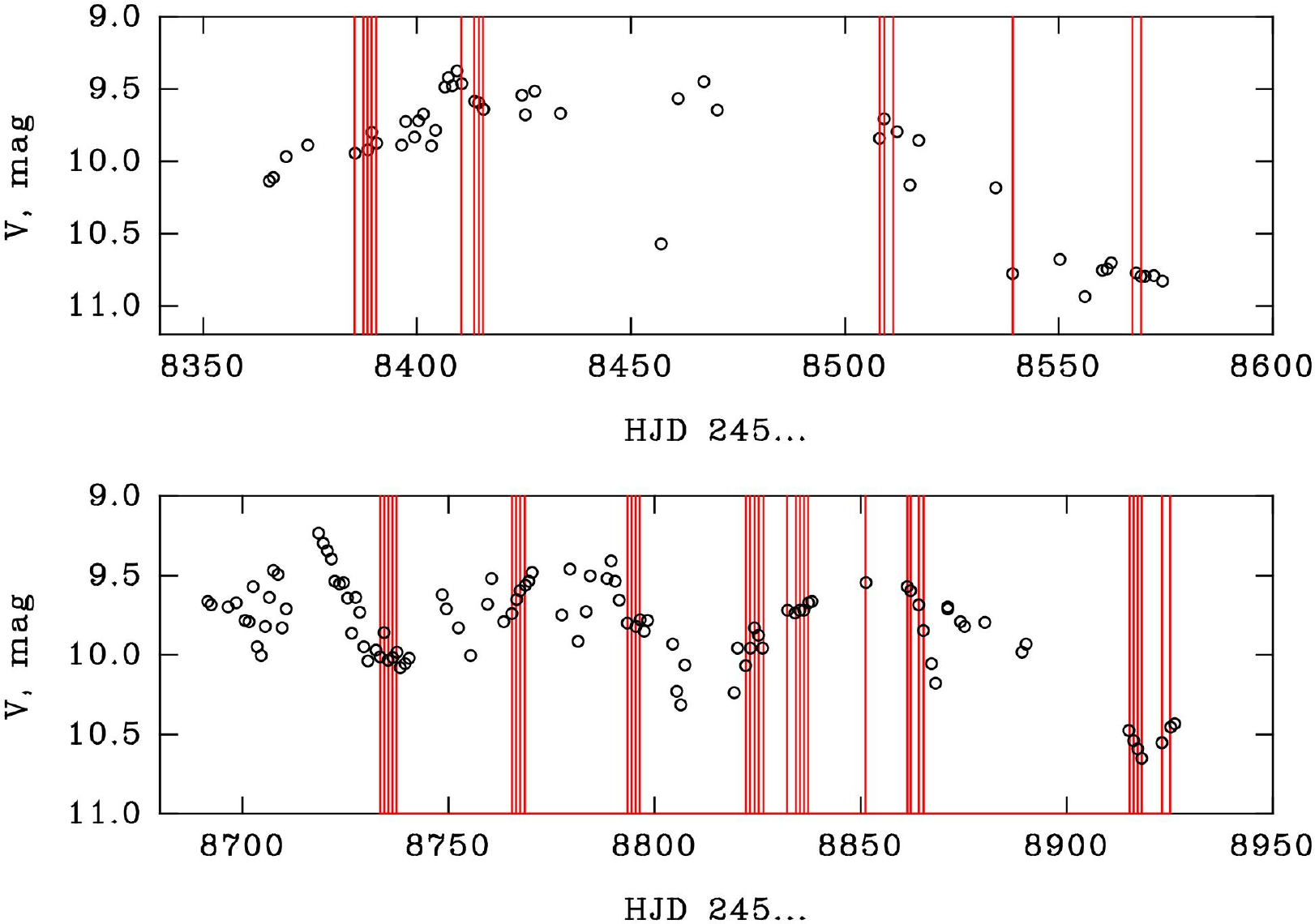}
    \caption{Light curve of RY Tau in two seasons, 2018-2019 (upper panel) and 2019-2020 (lower panel). Vertical lines mark the moments of spectral observations.}
    \label{fig:RY_V18}
\end{figure}

So far, we have collected over 160 nights of spectral observations of RY Tau during seven seasons, from 2013 to 2019. The light curves for the latest two seasons are presented in Fig.~\ref{fig:RY_V18}, where the moments of spectral observations are marked with vertical bars. The light curves and moments of spectral observations in the previous five seasons from 2013 to 2018 are given in  Paper I. 

\section{Results}

\begin{figure}  
\begin{minipage}{0.42\linewidth}
	\includegraphics[width=\linewidth]{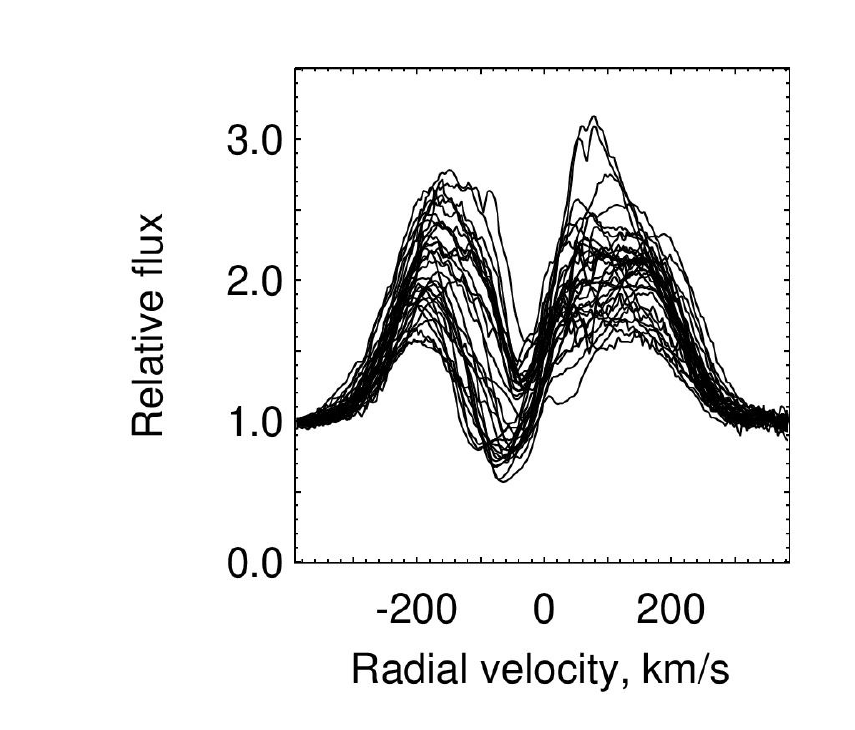}
\end{minipage}
\begin{minipage}{0.58\linewidth}
	\includegraphics[width=\linewidth]{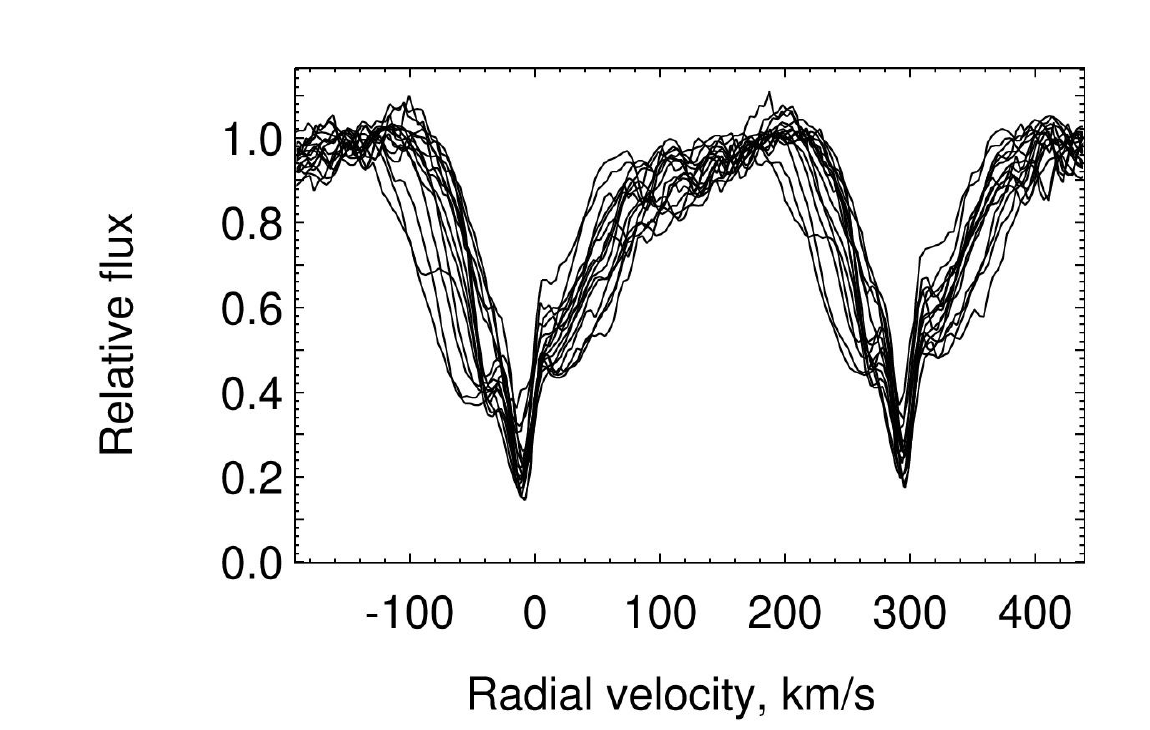}
\end{minipage}
    \caption{The ranges of \Hal emission (left) and \DNa absorption (right) profile variability in 2019-2020. In the region of \DNa lines the radial velocity scale is given for the D2 line.}
    \label{fig:fig2}
\end{figure}

\begin{figure}  
\center{\includegraphics[width={0.70\columnwidth}]{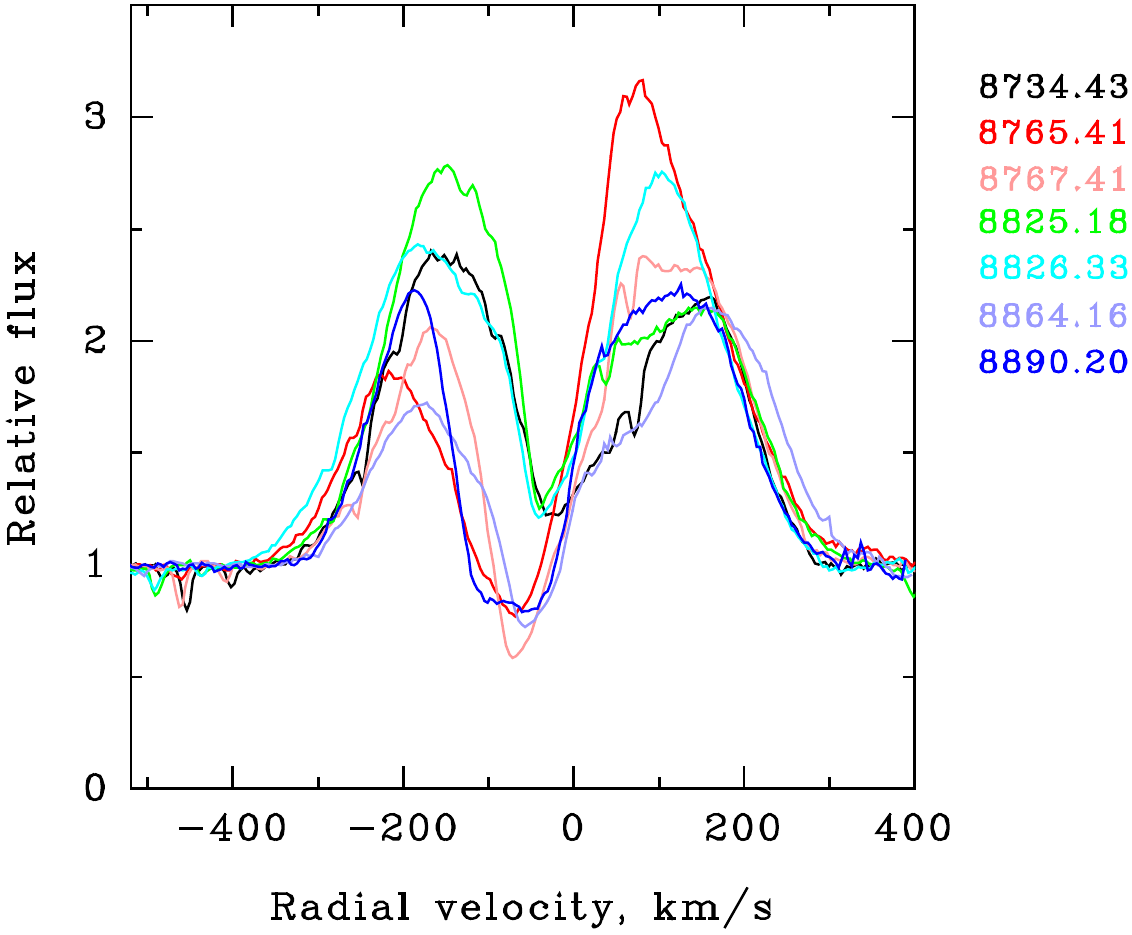}}
    \caption{Sample of  \Hal line profiles. Observation dates are displayed in the same colour as the corresponding profiles. }
    \label{fig:HA_PROF}
\end{figure}

\begin{figure}  
\center{\includegraphics[width={0.60\columnwidth}]{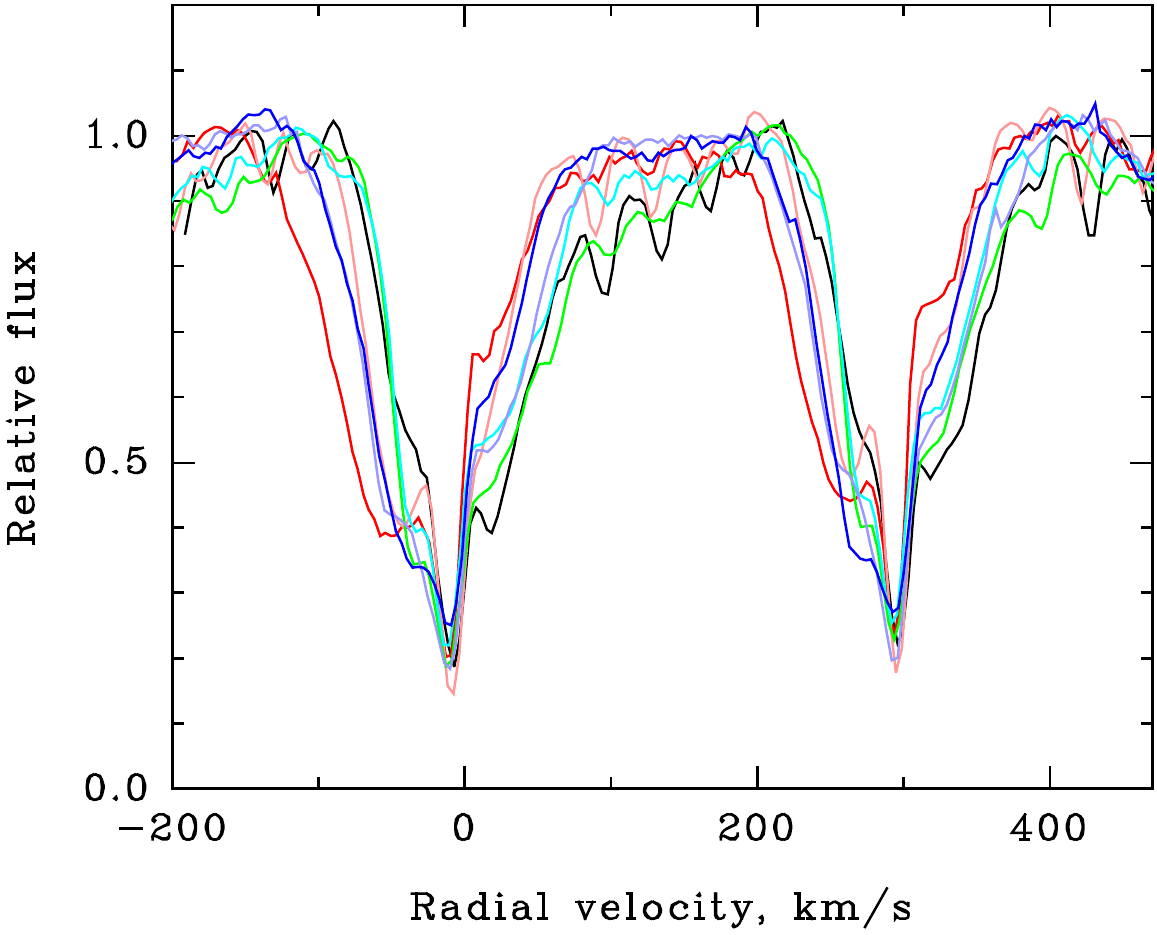}}
    \caption{ Sample of \DNa line profiles. Observation dates and colours are the same a in  Fig.~\ref{fig:HA_PROF}. }
    \label{fig:DNA_PROF}
\end{figure}

In the following, we analyse the \Hal and \DNa lines variability and interpret the results in terms of the accretion and wind gas flows.
The \Hal emission line has a broad profile with variable depression in the blue wing, sometimes below continuum, indicating absorption in the outflowing gas. The red wing is also variable in the intensity of emission.
The \DNa lines show a profile composed of several components: 1) photospheric absorption of a G2 type star, broadened by axial rotation to $v\sin i=52 \pm 2$ \kms, 2) narrow absorption dip of interstellar origin, 3) broad variable absorptions in the blue and red wings.
As an example, Fig.~\ref{fig:fig2} shows the ranges of \Hal and \DNa lines profile variability, observed in one season of 2019-2020. In this and the following diagrams, the radial velocity (RV) is astrocentric, assuming the heliocentric radial velocity of RY Tau is +18.0 \kms \citep{Petrov1999}.
The terms "blue" and "red" wing of a line correspond to negative and positive radial velocities.
Another example is shown in  Figs.~\ref{fig:HA_PROF} and ~\ref{fig:DNA_PROF}, where the observation dates are colour coded 
so the line changes over time can be traced.

Besides the intrinsic variability of \Hal  flux, the observed intensity of \Hal  emission depends on the circumstellar extinction: when the brightness of RY Tau drops, 
the equivalent width of \Hal emission gets higher (Paper I). 
This may be because circumstellar dust obscures the star, but not the whole area of \Hal emission. In this case, the total flux radiated in \Hal can be calculated as F = EW$\times 10^{-0.4 (V-10)}$, where the flux is expressed in units of the continuum flux density of a star with $V=10$ mag, which is $3.67 \times 10^{-13}$ erg cm$^{-2}$ s$^{-1}$ \AA$^{-1}$.  The photometric $R$ band would be  more appropriate for \Hal flux calibration, but for some moments of spectral observations only $V$ magnitudes are available.  In RY Tau the colour $(V-R)$ does not change considerably with brightness: on average, $(V-R) = 1.1 \pm 0.1$ mag. Therefore, the use of $V$ magnitude introduces a constant factor to the derived flux. 

\begin{figure}  
	\includegraphics[width=\columnwidth]{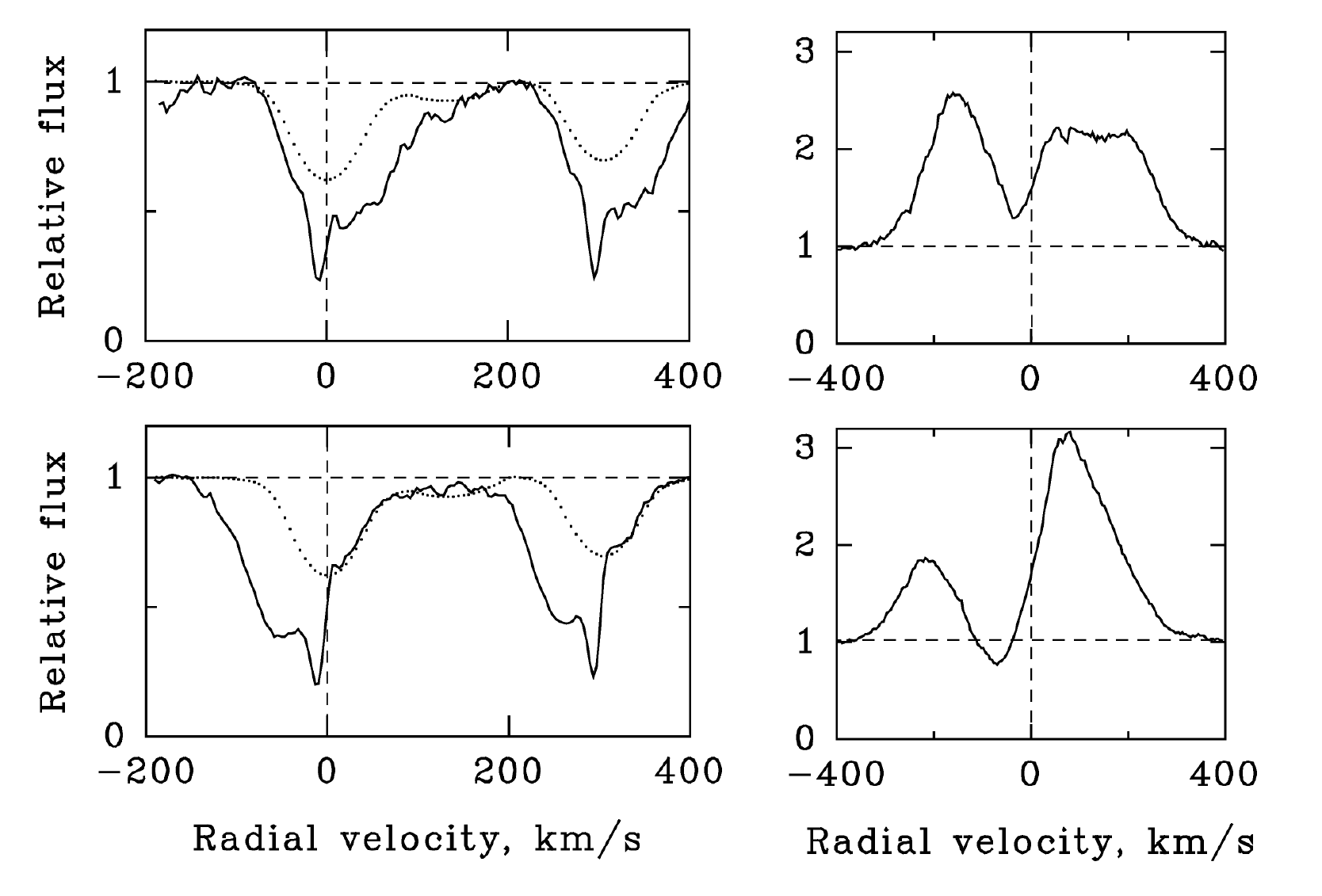}
    \caption{\DNa and  \Hal  line profiles in different dates, HJD =2458737.41 (upper) and 2458765.42 (lower). The photospheric components of the \DNa absorptions, broadened by stellar rotation to $v\sin i=52 $\kms, are shown as dotted profiles. The absorption dip at about zero velocity in \DNa profile belongs to interstellar absorption.}
    \label{fig:RY_AC_WI}
\end{figure}

The accretion and wind gas flows affect the red and blue wings of spectral lines, broadened by the Doppler effect. In \Hal the major changes are in the depth of blue-shifted absorption, while in the \DNa lines varying absorptions appear on the blue or red sides of the photospheric component. The most clear  example of  such a variability in \Hal  and \DNa lines is shown in Fig.~\ref{fig:RY_AC_WI}. 

In the \Na D2  line we measure equivalent widths of the blue and red wings, D2b and D2r, in radial velocity ranges [-160 ... -10] and [+10 ... +160] \kms correspondingly, i.e. avoiding the central absorption dip of circumstellar origin. The photospheric component of D2 contributes constant additions to the variable D2b and D2r quantities.

\begin{figure}  
\center{\includegraphics[width={0.7\columnwidth}]{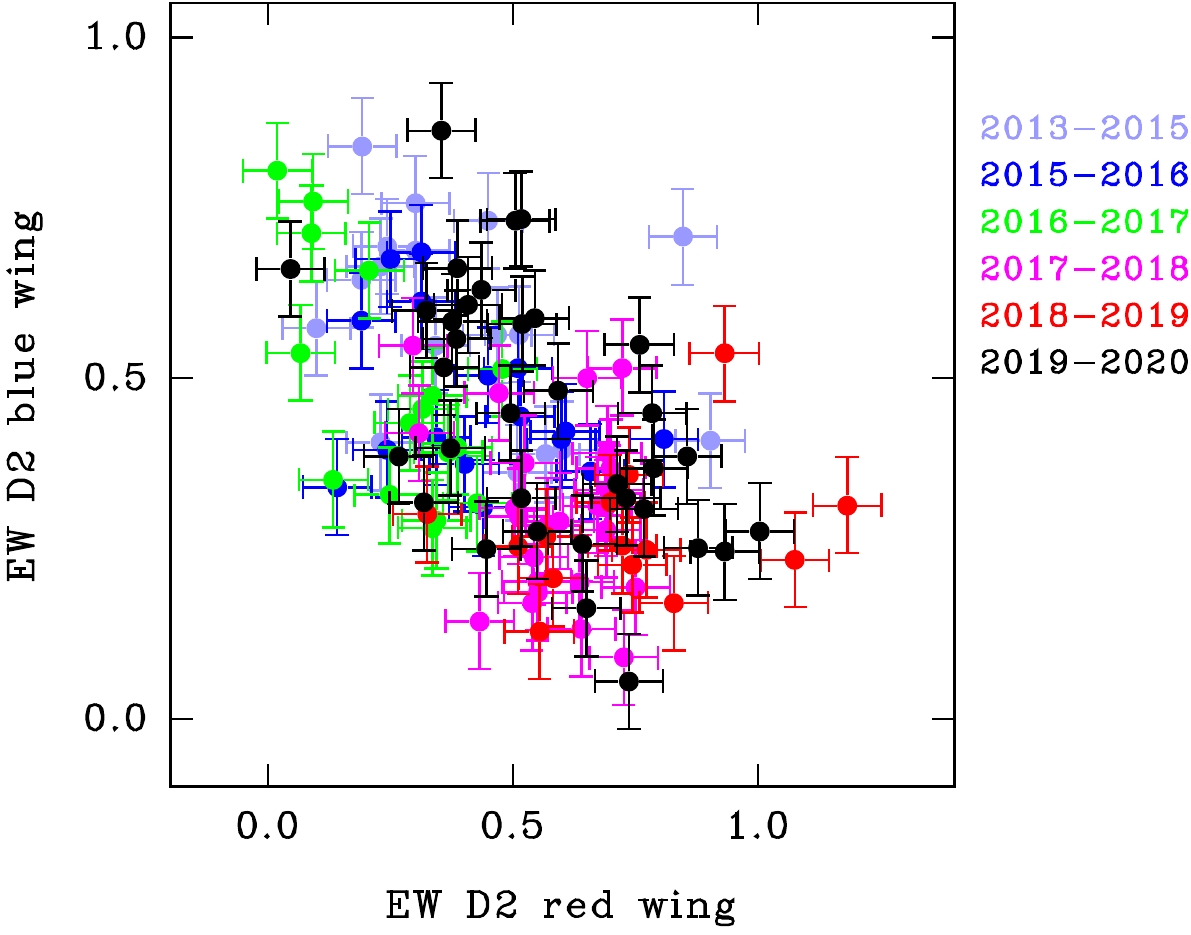}} 
    \caption{Inverse correlation between equivalent widths (EW, \AA) of the blue and red wings of D2 absorption. 
 The different seasons of observations are coded with colours.
}
\label{fig:DRB1320}
\end{figure}

\begin{figure}  
\includegraphics[width=\columnwidth]{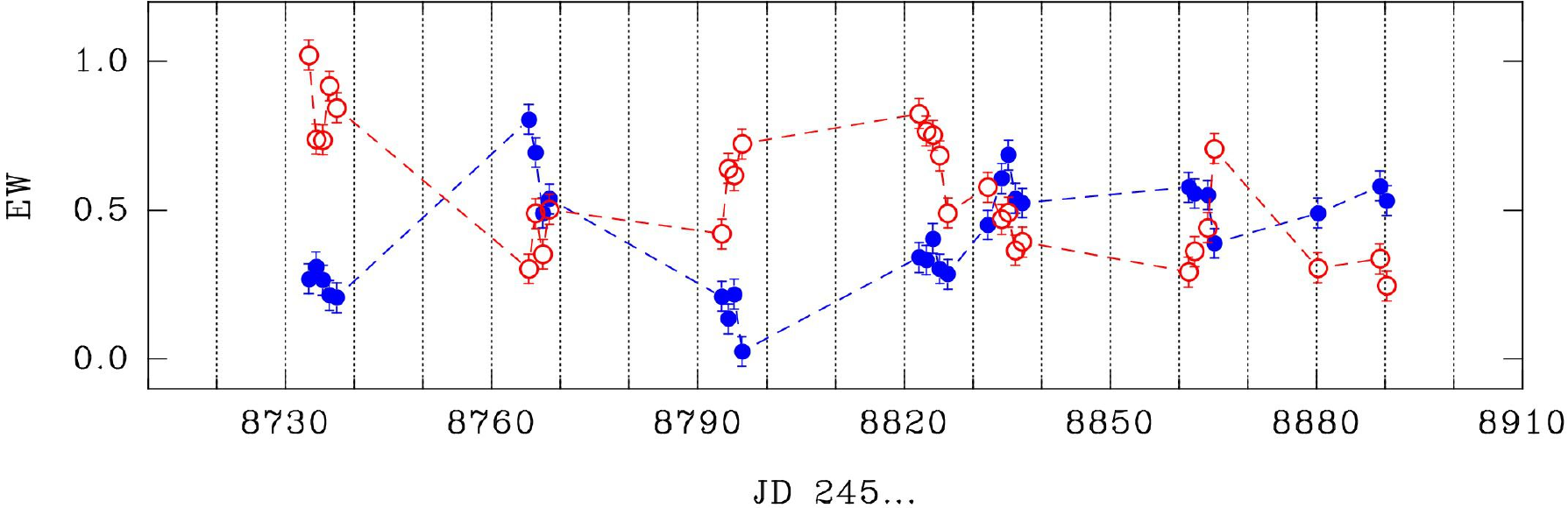}  
    \caption{Anti-phase variations of equivalent widths (EW, \AA) of the blue (filled circles) and red (open circles) wings of the D2 absorption during the season of 2019-2020.}
    \label{fig:D2_19-20}
\end{figure}

It was found that D2b and D2r vary in anti-phase: an increase in the blue-shifted absorption is followed by a decrease of the red-shifted absorption and vice versa. The inverse correlation between the two components is shown in Fig. ~\ref{fig:DRB1320}. 
 An example of time variability of D2b and D2r during one season of observations of 2019-2020 is shown in Fig. ~\ref{fig:D2_19-20}. 

\begin{figure}  
\center{\includegraphics[width={0.7\columnwidth}]{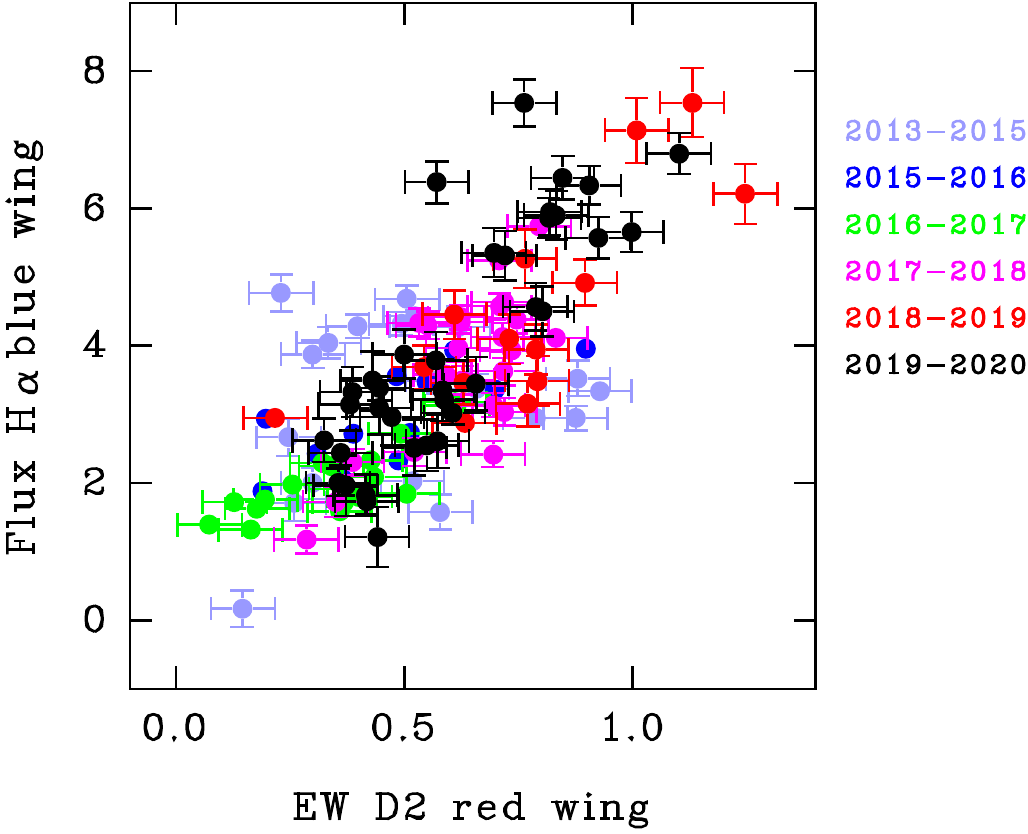}} 
    \caption{ Correlation in the variability of equivalent width (EW, \AA) of the red wing of Na\,{\sc i}\, D2\, absorption and the flux in the blue wing of \Hal emission.
The flux units:  $3.67 \times 10^{-13}$ erg cm$^{-2}$ s$^{-1}$. 
The different seasons of observations are coded with colours.}
    \label{fig:RY_HB_DR}
\end{figure}

In terms of gas flows, these results mean that either inflow or outflow prevails on the line of sight at any moment of observation.
Even more tight correlation appears in the variability of the red wing of \Na D2 absorption and the flux in the blue wing of \Hal emission (Fig.~\ref{fig:RY_HB_DR}). An increase of absorption in the red wing of \Na D2 is accompanied by an increase of emission in the blue wing of \Hal . The blue wing of \Hal is the best indicator of outflow: increase of emission flux in \Hal blue wing means fading of the wind. On the other hand, the increase of absorption in the red wing of D2 line signals the appearance of infalling gas on the line of sight. Therefore, the correlation in the Fig.~\ref{fig:RY_HB_DR} also indicates the inverse correlation between inflow and outflow on the line of sight. 

 The red wing of \Hal emission does not correlate with any part of the \Na D2 absorption.  

In the following, we analyse \textit{time variability} in \Hal and D2 line profiles.
The time series of \Hal profiles include 162 nights of observations in 2013-2010.
The \Hal emission is strong and easy to measure in each spectrum, while the D2 absorption profile in some spectra appears noisy, or contaminated by water lines, therefore the number of the \DNa spectra used in the analysis is less, about 120. 
The variables  are fluxes in \Hal emission line profile  and intensities in the \Na D2  line profile.
The Lomb-Scargle algorithm modified by \citet{Horne1986} was applied.

\begin{figure}  
	\includegraphics[width=\columnwidth]{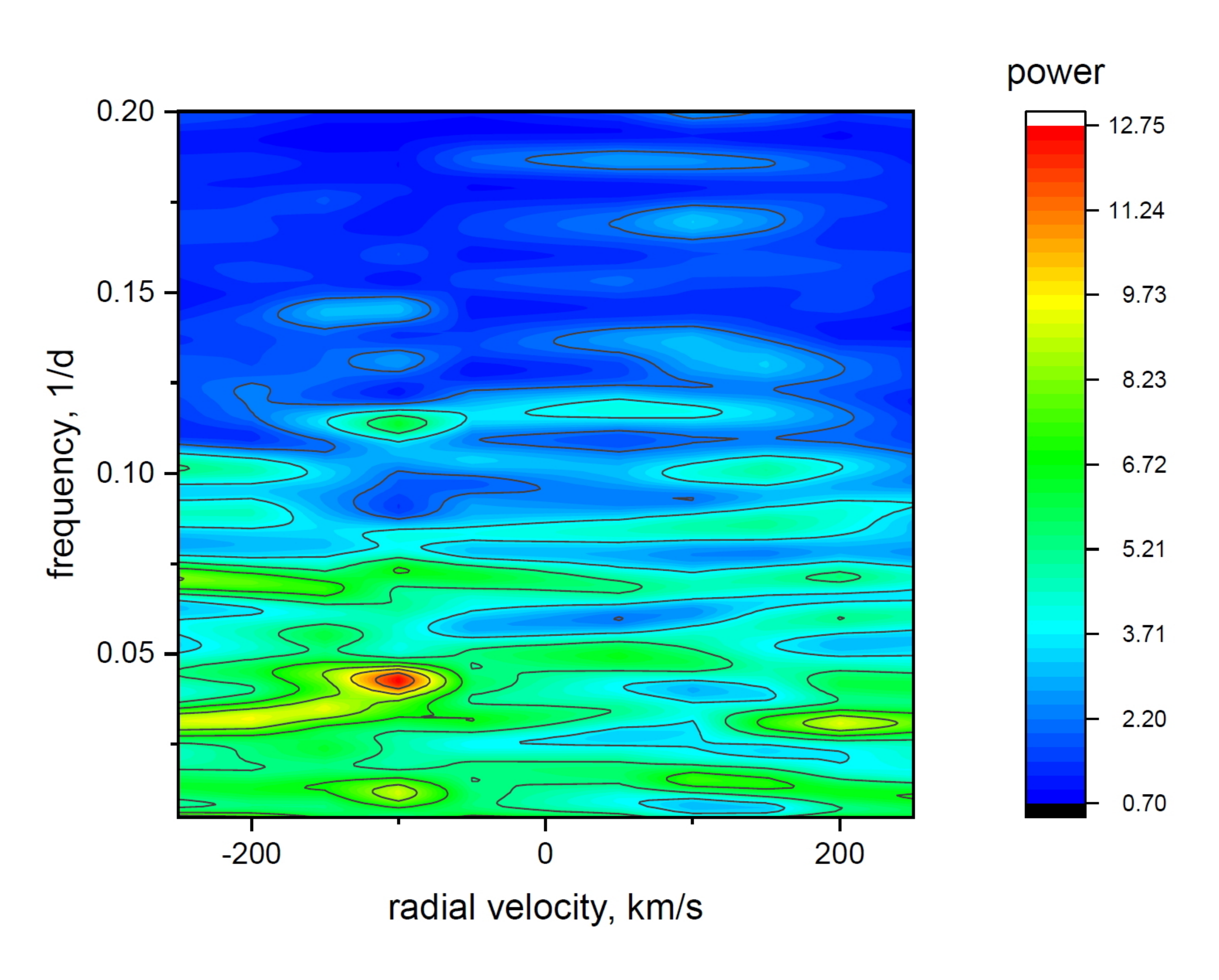}
  \caption{2D periodogram of \Hal flux. The maximum power is concentrated at radial velocity -100 \kms, at  frequency of about  0.04, corresponding to a period within 22-24 days }
    \label{fig:Ha_power}
\end{figure}

\begin{figure}  
\center{\includegraphics[width=0.7\columnwidth]{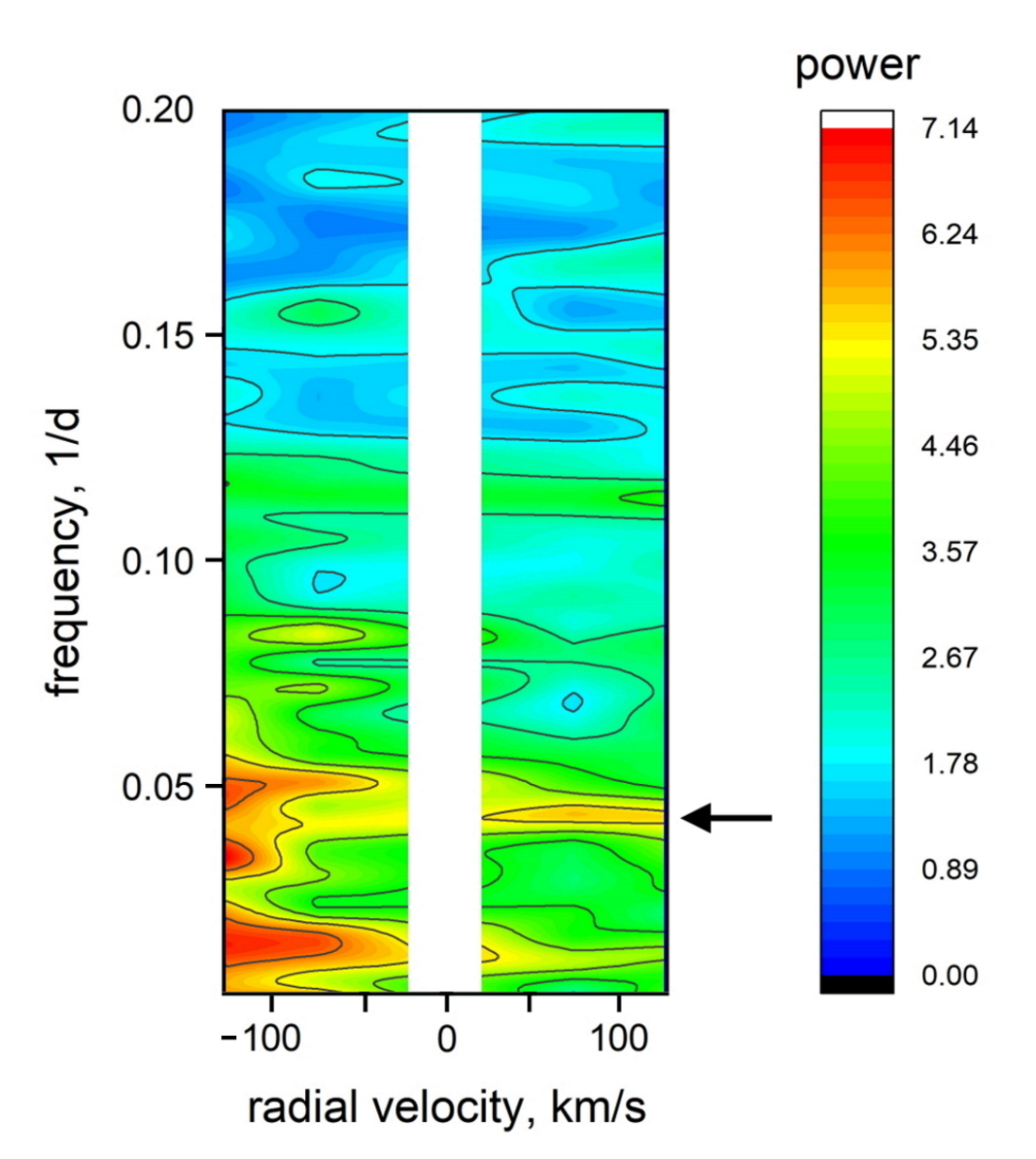}}
  \caption{2D periodogram of  \Na D2  intensity. The central blank strip is the excluded region of  interstellar absorption. Note the enhanced power in the red wing of the line at frequency of  0.044  (marked with arrow), corresponding to a  period of 22.7 days}
    \label{fig:d2_na_power_2}
\end{figure}

The Figs. ~\ref{fig:Ha_power} and ~\ref{fig:d2_na_power_2} show 2-D periodogram in the ranges of periods longer than 5 days and  the ranges
of radial velocities from -250 to +250 \kms in \Hal and from -125 to +125 \kms in \Na D2  line. 
In \Hal emission periodogram  there is a clear spot of high power at the radial velocity of  -100 km/s, at the frequency of about 0.04 day$^{-1}$,
corresponding to a period within  22-24 days. At the same frequency, the highest power is observed in the red wing of \Na D2  line.
Contrary to \Hal, the periodic modulation of the \Na D2  line is not confined to a narrow radial velocity range, but is observed
across the red wing of the line, from +50 to +150 km/s, at the frequency of 0.044 $\pm$ 0.002 day$^{-1}$, which corresponds to a period
of 22.7 $\pm$ 1.0 day.  In the blue wing of  \Na D2  line this period is blended with others in the range of 20-30 days.
Note that the absorption features measured in \DNa are relatively small and prone to errors due to the presence of the telluric water lines.

\begin{figure} 
\center{\includegraphics[width=0.8\linewidth]{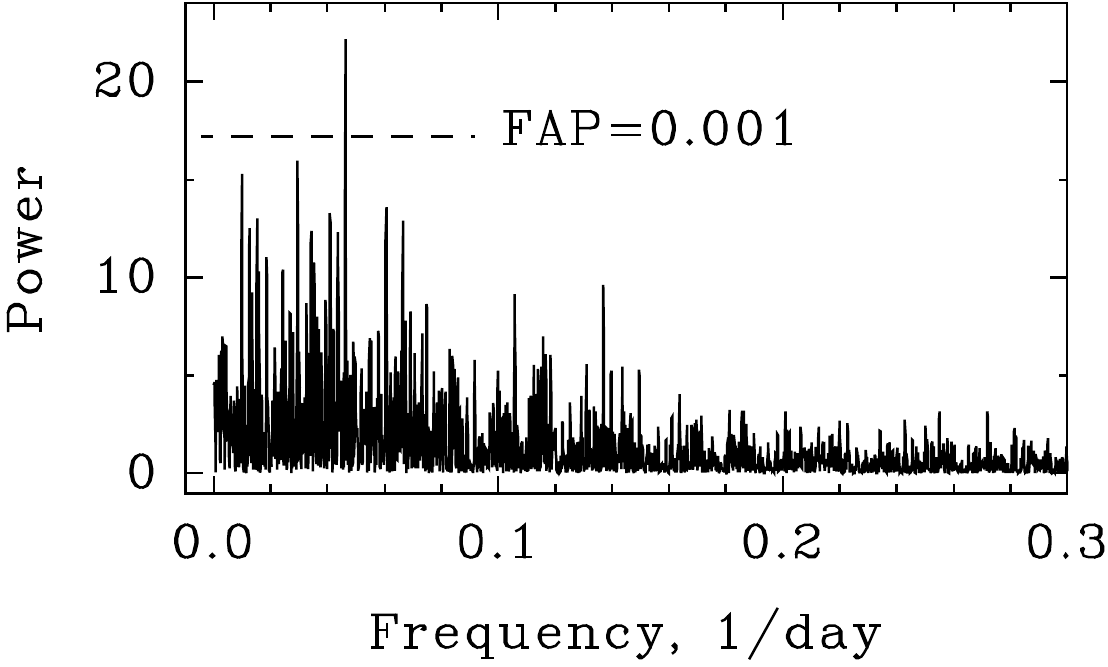}}
  \caption{Power spectrum of \Hal line asymmetry. The highest peak corresponds to the period of 21.6 days.}
    \label{fig:RY_POW}
\end{figure}

\begin{figure}  
\center{\includegraphics[width=0.75\linewidth]{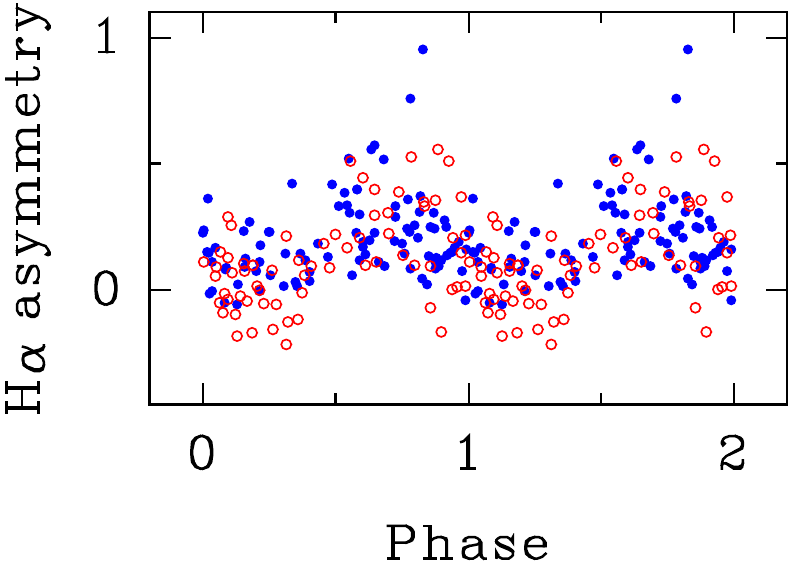}}
  \caption{\Hal line asymmetry convolved with  a period of 21.6 days. Asymmetry = 1 corresponds to the P Cyg type profile. Filled circles: data of 2013-2017. Open circles: data of 2017-2020.   The origin of time is the same in both samples: HJD=2456592.444  }
    \label{fig:RY_PHAS3}
\end{figure}

The periodic variations in \Hal  can be revealed also in analysis of the line profile asymmetry. 
As a line asymmetry, we use the difference-to-sum ratio (b-r)/(b+r), where 'b' and 'r' are equivalent widths of the blue and red halves 
of the line profile relative to zero radial velocity. Asymmetry = 1 corresponds to P Cyg profile, asymmetry = 0 corresponds to symmetric blue and red wings. 
The power spectrum of the \Hal line asymmety in the full set of our observations of RY Tau (seven seasons) reveals oscillations with the most probable period of 21.6 days, with false alarm probability FAP < 0.001 (see Fig.~\ref{fig:RY_POW}). 
The Fig.~\ref{fig:RY_PHAS3} shows the phase diagram of \Hal asymmetry. The data are split in two parts: 2013-2017 (93 dates of observations) and 2017-2020 (69 dates) to show that oscillations were present in both parts. 

The periodic variations in the blue wing of \Hal  occurs at radial velocity about -100 \kms, which is within the wide blue-shifted dip
associated with the wind. The Fig.~\ref{fig:HA_PROF}  shows that  radial velocity of -100 \kms falls within the steep slope of the dip, 
therefore the observed variations of line intensity at this part of the profile  may be caused by changes  in the outflow velocity. 

In the \DNa lines the variable absorption in the red wing is formed in the infalling gas.
The infall velocity range from +50 to +150 \kms covers almost the whole accretion channel within the magnetosphere. 
The oscillations in the red wing of the \DNa  mean that we observe {\it modulated accretion}.

\section{Discussion}

The emission line profiles broadened by the Doppler effect provide information about the main directions of gas flows -- accretion and wind \citep{Hartmann1998, Alencar2000}. The models of magnetospheric accretion and the disc wind were used to interpret the observed profiles of H\,{\sc i}\, and He\,{\sc i}\, (\citealt*{Kurosawa2006, Kurosawa2011, Kwan2011}; \citealt{Gahm2018}). 
Dynamics of wind in CTTS was studied in a number of spectral time series, e.g. 
\citet{Giampapa1993, Johns1995}, \citet*{Alencar2001}, \citet{Alencar2005,Takami2016, Petrov2019}.

Periodic variations in the accretion and wind features were first detected in spectrum of SU Aur \citep{Giampapa1993, Johns1995}
-- a fast rotating TTS  with stellar parameters  resembling those of RY Tau.
\citet{Johns1995} suggested a model of oblique rotator (the 'eggbeater' model)  to explain how the  wind 
and the accretion flows replace each other on the line of sight as the star rotates.
More advanced models  were elaborated later to explain the observed periodic modulations in the photometric and spectroscopic features in CTTS. 
If the large-scale stellar magnetosphere is inclined onto the stellar rotational axis, interaction of magnetic field
with the inner disk results in the formation of a strong bending wave (a warp).
The warp is close to the radius of corotation and  rotates with the angular velocity of the star, like the warp observed in 
AA Tau \citep{Bouvier1999, Bouvier2013}, and the warps observed in the global 3D simulations \citep{Romanova2013}. 

In case of  RY Tau the observed 22-days period is much longer than the 3-days  period of stellar rotation, therefore the the model of inclined magnetosphere  cannot be applied. 
It is for the first time the  periodic switches of the accretion and wind flows, not related to stellar rotation,  is observed in a CTTS. An interpretation of this phenomenon should answer the question: is this an intrinsic property of the magnetospheric accretion process, or is it induced by an external force, e.g. orbital motion of a planet or a substellar companion?

A powerful instrument for analysis of the gas flows around a young star is the numerical simulation of the accretion and winds \citep[see review by][]{Romanova2015}. Formulation of the basic concepts can be found in \citet{Camenzind1990,Shu1994}; \citet*{Lovelace1995}.

In CTTS the accretion and winds are physically related processes, where the winds are consequences of accretion, either the magnetospheric accretion or the disc accretion. Although the magnetic fields of CTTSs are multipolar, it is the dipole mode that truncates the accretion disc and controls the magnetospheric accretion \citep{Johnstone2014}.

The main parameters that determine the modes of accretion and wind are stellar mass M, the angular velocity of rotation of the star $\Omega$, the magnetic field of the star (the magnetic moment of the dipole $\mu$), and the mass accretion rate $\dot {\textrm M}$. With these parameters, the corotation radius R$_{co}$ and the radius of stellar magnetosphere R$_m$ are expressed as: R$_{co}^3 =$ G M / $\Omega^2$, and R$_m \approx \mu^{4/7}  / (\dot {\textrm M^2}$ G M)$^{1/7}$ \citep{Koenigl1991}. The condition R$_{co} = $ R$_m$ defines the boundary between the regime of accretion and the regime of propeller \citep{Romanova2015}.

In case of the fast rotating RY Tau, R$_{co} \approx 3$ R$_{star}$. The magnetospheric radius of a CTTS is typically about 7 stellar radii \citep{Bouvier2007}, although may be within 3 to 10 in individual stars \citep{Johnstone2014}.  RY Tau is most probably in the propeller regime. The presence of the extended jet in RY Tau is an evidence in favor of a strong propeller.

In the propeller regime, the condition R$_m$ > R$_{co}$ does not prevent the accretion completely, but makes it non-permanent. The centrifugal barrier prevents the accretion of matter in the equatorial plane. However, the matter may flow above or below the closed parts of the magnetosphere and accrete onto a star from poles.

Axisymmetric simulations show the presence of both accretion and outflows during the propeller regime \citep[e.g.][]{Romanova2005,Romanova2009,Ustyugova2006}. Matter accretes inward, accumulates in the inner disc and diffusively penetrates through the external layers of the closed magnetosphere. When some critical amount of matter accumulated inside the magnetosphere and the centrifugal force  becomes larger than the magnetic tension force, the closed field lines inflate, and the accumulated matter expells in the wind. At the same time matter accretes onto a star from the opposite side of the disc, where the magnetosphere is closed \citep{Lii2014,Romanova2018}. The process is cyclic. After unloading matter, the inner disc becomes 'lighter', the magnetosphere expands and pushes the inner disc outward. After that the process of accumulation/diffusion repeats.
\footnote{Cyclic accretion and ejections were observed by \citet{Goodson1997, Zanni2007} in a non-propeller regime, where the cycle is connected  with periodic inflation of the field lines \citep[see also][]{Lovelace2010}.}

The time-scale and the level of periodicity between the accretion/ejections events depends on several factors, including the viscosity (accretion rate) in the disc and  diffusivity at the disc-magnetosphere boundary. For example, at small values of diffusivity $\alpha=0.02$ \citep[used in][]{Romanova2018}, events of 
accretion/ejections are episodic, and only occasionally quasi-periodic. However, in experimental simulations of accretion at higher values of diffusivity, $\alpha_d=0.2$ and viscosity $\alpha_v=0.6$ in the disc, these events become almost periodic \citep[see Fig. 3 from][]{Romanova2005}. \citet{Goodson1997} and \citet{Zanni2009} also used relatively high viscosity and diffusivity ($\alpha_d=0.2$ and 1.0, respectively), and observed almost periodic events of inflation of the magnetic field lines and ejections to the wind in non-propelling stars. All above simulations are axisymmetric, and exclude important instabilities at the disc-magnetosphere boundary, such as Rayleigh--Taylor instability. 3D simulations of accretion in non-propeller regime show that this instability provides high efficient diffusivity at the disc-magnetosphere boundary \citep[e.g.][]{Romanova2015}. We suggest that accretion/ejection events could be more periodic in more realistic case of higher diffusivity.  

In application to RY Tau, the axisymmetric simulations show episodic or quasi-periodic ejections with time-scales of 10-100 days, depending on the parameters of the model. Global 3D MHD simulations of the propeller regime are required to understand whether these ejections can be periodic.
 Magnetic fields of some T Tauri stars are known to change on a time scale of years, (e.g. \citet{Donati2011}, \citet{Yu2019} ), therefore the characteristic time of the quasi-periodic ejections  may be slowly variable. 

Another characteristic of the propeller is that the direction of the accretion funnel (and outflows) switches between upward and downward relative to the disc plane \citep{Lovelace2010}. In this case, on the line of sight to the star one should observe alternation of the wind and accretion features. It seems to be a reasonable explanation of the observed 'flip-flops' of accretion and wind flows in RY Tau, but this scenario is not perfect either. Due to the inherent turbulence of the gas flows, the actual direction of the accretion funnel (above or below the disc plane) in each cycle cannot be predicted. The numerical simulations show that the path of the accretion is not necessarily changed in each cycle: a new cycle can start with the same directions, and the oscillation phase is not preserved. In this case, one cannot expect the phase diagram like that in Fig.~\ref{fig:RY_PHAS3}.

Alternatively, the stable cycling of the accretion/ejections events may be induced by a planet orbiting near the inner edge of the accretion disc. This scenario has not been studied in detail by numerical simulations, except for the research by \citet{Teyssandier2020}, where it was first shown how the accretion rate at the inner disc edge is regulated by a massive planet at excentric orbit.


A period of about 23 days in variations of emission line intensities in UV and optical spectrum of RY Tau was reported earlier by \citet{Ismailov2011} and interpreted as a consequence of a protoplanet orbiting in the inner accretion disc. 
The presence of a planet or a sub-stellar companion to RY Tau has been suggested also from the fact that jet of the star is wiggled. \citet{Garufi2019} estimated that the observed amount of the jet wiggle is consistent with the presence of a precessing disc warp or misaligned inner disc that would be induced by a gravitating object -- a giant planet or a sub-stellar companion. 
The relation between the mass and orbital distance of a putative planet responsible for the jet wiggle in RY Tau \citep[Fig. 6 in][]{Garufi2019} indicates that it may be a planet with mass 3-4 M$_J$ at the distance of about 0.2 AU.

One may expect that the disk warp or inclined inner disk could be noticed from the photometric series of the star. 
The NIR interferometry of RY Tau showed that the star is occulted by the disc surface layers close to the sublimation rim \citep{Davies2020}. 
A similar conclusion was made from the photometric data: the star is permanently obscured by the dusty disc wind, 
with the minimal circumstellar extinction Av =1.6 mag (Paper I). 
Analysis of  long photometric series of RY Tau \citep{Artemenko2010} did not reveal any stable period around  22 days,  although  20 days period was
noticed earlier in photometric observation of 1993 \citep{Zajtseva2010}. 
Probably,  the distortions of the inner disk, associated with the putative planet, are  mostly within the sublimation radius, in a dust-free zone.

With the mass of RY Tau 2.08 \Msun, the period of 22 days corresponds to Keplerian orbit at 0.2 AU. The inclination angle of RY Tau ($i = 65^\circ$) enables to confirm the planet with the radial velocity technique: expected amplitude is $\geq 90$ m s$^{-1}$ if a planet mass is $\geq 2$ M$_J$.

\section{Conclusions}

The seven years monitoring of RY Tau revealed a highly probable period of $\sim$22 days in variations of spectroscopic signatures of accretion and wind in \Hal and \DNa line profiles.  The absorptions in the infalling and the outflowing gas streams vary in anti-phase: increase of  infall is accompanied by decrease  of  outflow, and vice versa.
The found period is much longer than that of the axial rotation of the star. If Keplerian, the period corresponds 
 to a distance of 0.2 AU, which  is close to the dust sublimation radius in this star. 
Although the oscillations of the accretion and wind flows may be explained by the MHD processes at the  disk-magnetosphere boundary in the propeller regime,
the phase stability of the observed modulations  suggests the effect of an external force.
As a tentative interpretation, we suggest a gravitational effect of a planet  orbiting at 0.2 AU. 

The presence of the putative planet may be confirmed by radial velocity measurements: expected amplitude is  $\geq 90$ m s$^{-1}$ if a planet mass is $\geq 2$ M$_J$.
Alternatively, if a more extended series of observations reveals a change in the modulation period, the planet hypothesis will be ruled out.

\section*{Acknowledgements}

The Crimean observations in 2019-2020, including data processing and data analysis by S. Artemenko, E. Babina, and P. Petrov, were supported by a grant from the Russian Science Foundation 19-72-10063.
M.M. Romanova acknowledges the NSF grant AST-2009820.
K.N. Grankin acknowledges the partial support from the Ministry of Science and Higher Education
of the Russian Federation (grant 075-15-2020-780).
S.Yu.Gorda was supported in part by the Ministry of Science and Higher Education of the Russian Federation within the framework of the research activities (project no. FEUZ-2020-0030).

\section*{Data availability statement}

The data underlying this article will be shared on reasonable request to the corresponding author [PP].



\bibliographystyle{mnras}







\bsp	
\label{lastpage}
\end{document}